\documentclass[%
 reprint,
 amsmath,amssymb,
 aps,
]{revtex4-2}

\usepackage{graphicx}
\usepackage{dcolumn}
\usepackage{bm}

\usepackage{dsfont}

\begin{document}

\title{Recursive Work Extraction from Quantum Conditional Information}

\author{Daegene Song}
\email{email: dsong@cbnu.ac.kr}
 \affiliation{Department of Management Information Systems, Chungbuk National University, Cheongju, 28644, Republic of Korea.}

\date{\today}

\begin{abstract}

Quantum superposition, a cornerstone of quantum mechanics, enables systems to exist in multiple states simultaneously, giving rise to probabilistic outcomes. In quantum information science, conditional entropy has become a key metric for quantifying uncertainty in one system given information about another, revealing non-classical correlations that transcend classical physics. This study examines the nature of quantum conditional entropy and reports two key findings. First, it demonstrates that probabilistic outcomes involving quantum superposition arise from work based on information about the eigenstate in a recursive process. Second, it proposes that this extractable work constitutes the energy available to living systems—a concept without a classical analogue—counteracting the natural tendency toward disorder.

\end{abstract}

\maketitle


\section{Introduction}
A qubit, or quantum bit, represents the fundamental unit of quantum information, analogous to a classical bit but with the key distinction of existing in a superposition of states \cite{QC1,ladd}. While classical bits are restricted to binary values of 0 or 1, a qubit can be in a superposition, described as \( a|0\rangle + b|1\rangle \). Upon measurement, the qubit yields a definite state, yielding 0 with a probability of \( |a|^2 \) and 1 with a probability of \( |b|^2 \).

The superposition property, combined with the probabilistic nature, underpins a wide range of groundbreaking applications. Superposition enables entangled particles to exhibit nonlocal correlations that defy classical physics \cite{bell}, even when separated by vast distances \cite{aspect,tittel}, challenging the principle of locality \cite{epr}. In quantum computing \cite{nielsen,QC2}, qubits use superposition to process multiple computational paths simultaneously, leading to computational speeds that far surpass those of classical systems \cite{deutsch,jozsa,shor,preskill}. Additionally, quantum cryptography exploits superposition to securely encode and transmit information \cite{bb84,comm}.

This research builds upon recent developments in quantum information science \cite{wilde,hayashi} to investigate the probabilistic characteristics of quantum systems. Conditional entropy, a key concept in information theory \cite{shannon,cover}, measures the uncertainty of one variable given knowledge of another. In the quantum realm, conditional entropy exhibits distinct non-classical behavior, including the potential for negative values \cite{cerf,horodecki}. The study examines how work can be extracted from conditional entropy \cite{rio} and applied within subsystems, influencing probabilistic outcomes through a recursive process. Importantly, this extractable work corresponds to energy—a resource harnessed by living systems, representing a phenomenon with no classical analogue. The result aligns with the observation that living systems maintain low entropy by continuously exchanging energy and matter with their environment, which allows them to stay organized and function despite the natural tendency toward disorder \cite{schrodinger,miller,schneider,kleidon}.

\section{The Setting}

\begin{figure}
\begin{center}
{\includegraphics[scale=.4]{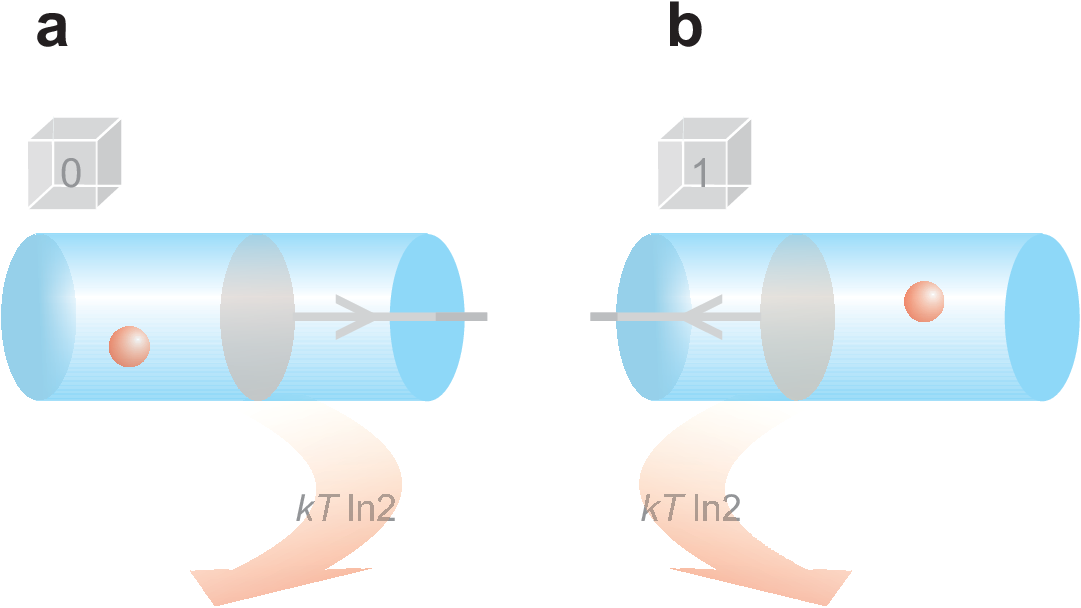}}
\end{center}
\caption{ {\bf{Work extraction based on the given information.}} Szilard's engine is a thermodynamic thought experiment involving a single gas particle in a volume, where measuring the particle's position enables the extraction of work. Based on the information of one bit, an amount of work equal to \(kT\ln 2\) can be extracted by selecting the appropriate side, as shown in {\bf{a}} and {\bf{b}}.} 
\label{extract}\end{figure}

A convenient way to express the basic unit of quantum information, i.e., a qubit, \( |\psi\rangle = a|0\rangle + b|1\rangle \), is through the Bloch sphere representation: \(  |\psi\rangle\langle\psi| = \frac{1}{2}(\mathds{1} + \hat{\mu} \cdot \vec{\sigma}) \), where \( \vec{\sigma} = (\sigma_x, \sigma_y, \sigma_z) \) are the Pauli matrices, and \( \hat{\mu} \) is a unit vector in the direction \( (\alpha, \beta) \) on the Bloch sphere, with \( 0 \leq \alpha \leq \pi \) and \( 0 \leq \beta < 2\pi \). To measure a qubit, an observable is used to provide a reference frame, which can be described by a unit vector \( \hat{v} = (\sin\gamma \cos\delta, \sin\gamma \sin\delta, \cos\gamma) \), with the corresponding operator expressed as \( \hat{v} \cdot \vec{\sigma} = |\hat{\omega}\rangle\langle\hat{\omega}| - |\hat{\omega}^{\perp}\rangle\langle\hat{\omega}^{\perp}| \). Here, \( |\hat{\omega}\rangle \) and \( |\hat{\omega}^{\perp}\rangle \) are the orthogonal eigenstates corresponding to the chosen basis, with \( |\hat{\omega}\rangle = \cos(\gamma/2) |0\rangle + e^{i\delta} \sin(\gamma/2) |1\rangle \). In particular, the measurement direction corresponds to the vector \( \hat{v} \), which points in the \( (\gamma, \delta) \) direction on the Bloch sphere, where \( 0 \leq \gamma \leq \pi \) and \( 0 \leq \delta < 2\pi \).

Now, consider introducing an ancillary system \( A \), defined as \( A \equiv A_1 \otimes A_2 \otimes A_3 \cdots A_N \) for some finite \( N \), where each subsystem is initially in the zero state. To examine the measurement process \cite{neumann}, the initial state of the qubit \( Q \) and the ancilla \( A \) can be written as \( |\phi_0\rangle_Q \otimes |0\rangle_{A_1} \otimes |0\rangle_{A_2} \cdots |0\rangle_{A_N} \). Suppose the observable \( \hat{v} \cdot \vec{\sigma} \) is chosen to measure the system \( Q \), establishing a correlation between the qubit and the ancilla via an interaction Hamiltonian. This results in the state \( \left( c_0 |\hat{\omega}\rangle_Q |0\rangle_{A_1} + c_1 |\hat{\omega}^{\perp}\rangle_Q |1\rangle_{A_1} \right) \otimes |0\rangle_{A_2} \cdots |0\rangle_{A_N} \). This correlation encodes the basis in the qubit \( Q \) and links it to system \( A_1 \). After tracing out \( Q \), the outcome for \( A_1 \) is either \( |0\rangle_{A_1} \) or \( |1\rangle_{A_1} \), which are directly observable, occurring with probabilities \( |c_0|^2 = |\langle \hat{\omega}|\phi_0\rangle|^2 \) and \( |c_1|^2 = |\langle \hat{\omega}^{\perp}|\phi_0\rangle|^2 \), respectively. For simplicity, the case where the probabilities for each outcome are equal, i.e., \( |c_0|^2 = |c_1|^2 = \frac{1}{2} \), will be considered for the remainder of this paper.

At this stage, while \( A_1 \) registers the result, the remaining ancillary systems \( A_i \) for \( 2 \leq i \leq N \) remain in the zero state. Consider a process in which the outcome stored in \( A_1 \) is sequentially copied to \( A_2 \), then from \( A_2 \) to \( A_3 \), and so forth, up to \( A_N \), such that the entire system \( A \) eventually contains the single bit of information originally recorded in \( A_1 \). Note that the classical correlations within the ancillary system \( A \), created through this copying process from \( A_1 \) to subsequent systems, could also arise from multipartite entanglement between the qubit \( Q \) and the ancillary system \( A \) from the outset \cite{cerf1}.

\section{Work Extraction and Conditional Information}

\begin{figure}
\begin{center}
\includegraphics[width=.4\textwidth]{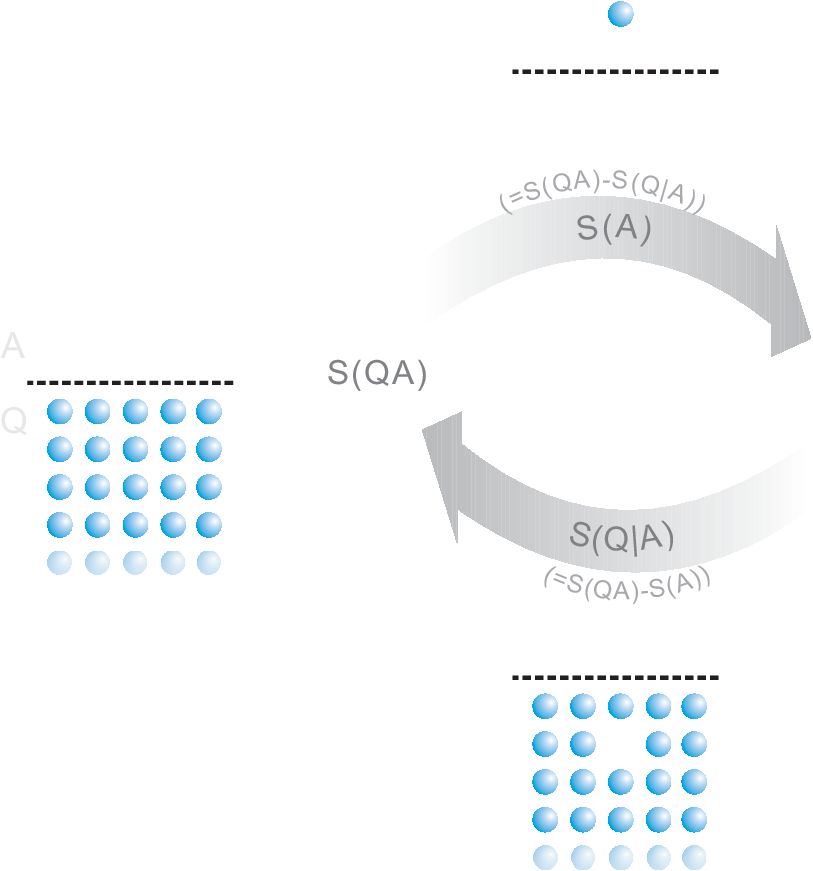}

\end{center}
\caption{ {\bf{Quantum conditional entropy and the antiqubit in the information sea.}} When systems \( Q \) and \( A \) are correlated, the quantum conditional entropy \( S(Q|A) \) can become negative, indicating a reversal of information about \( Q \) relative to \( A \). This process can be visualized using Dirac sea notation, where the vacuum is imagined as being filled with negative information. The information about \( Q \), relative to the outcome of system \( A \), can be interpreted as an antiqubit or a hole in the information sea. }
\label{diracsea}
\end{figure}

Landauer’s principle outlines a connection between seemingly abstract information and tangible physical quantities, particularly in the context of energy and thermodynamics \cite{landauer1}. Erasing a single bit of information incurs a fundamental energy cost of \( kT\ln 2 \), where $k$ is Boltzmann's constant and $T$ is the temperature of an environment, which leads to an increase in entropy and the release of heat, thus linking information processing to physical phenomena \cite{landauer,bennett,vedral1,berut,goold}.  Moreover, Landauer's principle implies that work can be extracted when information about a system is obtained \cite{rio}. A classical example is Szilard's engine: in a closed volume, where a particle may occupy the left (denoted \(0\)) or right (\(1\)) side \cite{szilard}, knowing the particle’s position allows work to be extracted through isothermal expansion, such as by attaching a weight to the appropriate side (Figure \ref{extract}).

This conversion of information into usable work can also be extended to quantum systems. If a system starts in a known state and is allowed to become randomized, the increase in entropy can be harnessed to perform work. One illustrative example \cite{lubkin,rio} involves a qubit initially in the ground energy state \( |0\rangle \), a pure state. When the qubit is coupled to a thermal bath at temperature \( T \), and through an infinitesimal perturbation, the probability of finding it in the excited state \( |1\rangle \) is given by \( p_1 = \frac{e^{-E/kT}}{1 + e^{-E/kT}} \). In the high-energy limit, the extracted work can be computed as \( \int p_1 \, dE = kT \ln 2 \), while the initial state transforms into a completely mixed state.

Therefore, just as in the classical case, information about the qubit's state enables the extraction of work equal to \( kT \ln 2 \). Applying this relationship between work and information to systems \(Q\) and \(A\), outlined earlier, the following conclusion can be drawn:
\begin{enumerate}
\item [(L1)] Knowing the outcome of \(A\) and the choice of basis \(\hat{v}\), an amount of work equal to \(kT \ln 2\) can be extracted from the system \(Q\).
\end{enumerate}
In other words, given the outcome of \(A\), the state of \(Q\) can be identified as one of the eigenstates, \( |\omega\rangle_Q \) or \( |\omega^{\perp}\rangle_Q \), which then enables the extraction of work.

\begin{figure}
\begin{center}
{\includegraphics[scale=.4]{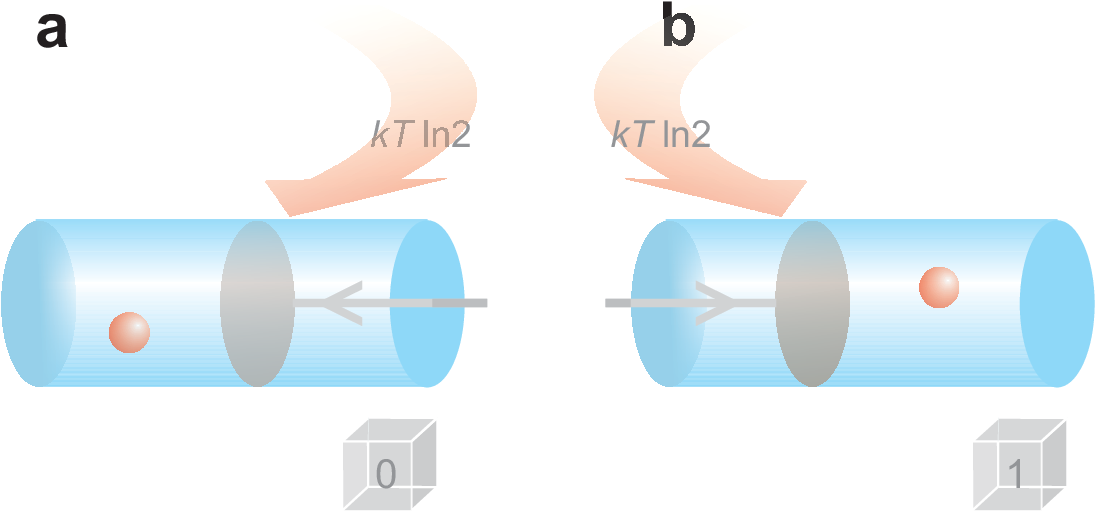}}
\end{center}
\caption{ {\bf{Doing work to reset to the left or right.}} Landauer's principle illustrates the connection between information and physical systems, showing that information processing is governed by thermodynamic limits. Resetting the particle's position to a known state, with the particle confined to either the left ({\bf{a}}) or right ({\bf{b}}) side of the volume, requires an amount of work equal to \(kT \ln 2\). Erasing to the right side can also be carried out by first resetting it to the left side of the volume, then interchanging it with the right side. } 
\label{do}\end{figure}

Conditional entropy measures the remaining uncertainty in one random variable given knowledge of another, indicating how much additional information is needed to describe it \cite{shannon,cover}. In quantum systems, this concept generalizes to quantum conditional entropy, expressed as \( S(Q|A) = S(QA) - S(A) \) \cite{cerf,horodecki,cerf1}, which quantifies the uncertainty of one quantum system relative to another. Unlike classical entropy, quantum conditional entropy can take negative values, a phenomenon with no classical counterpart, as correlations between systems reduce uncertainty in ways that exceed classical limits.

Dirac outlined a theoretical concept of a sea in which all negative energy states in a quantum vacuum are filled, thus preventing electrons from occupying these states \cite{dirac}. When an electron is excited out of this sea, it leaves behind a hole that behaves like a positively charged particle, leading to the prediction and discovery of the positron, the electron's antiparticle. In a similar vein, negative conditional entropy can be viewed as a sea of negative information. In particular, authors in \cite{cerf} introduced the notion of an antiqubit in relation to negative entropy, where a quantum state, like an antiparticle, cancels out information when paired with a corresponding qubit. Here, the conditional entropy \(S(Q|A)\) represents the negative information that the qubit \(Q\) possesses relative to the result of \(A\). In this framework, the quantum system that encodes the basis choice behaves as if it moves backward in time, functioning as an antiqubit with respect to system \(A\) (Figure \ref{diracsea}).  This makes it causally plausible to infer the outcome of \(A\) based on the conditional entropy \(S(Q|A)\).

The reversal process of work extraction can be described as follows: when the position of the particle within the volume is unknown, the work required to reset it to either the left (denoted as 0) or the right (denoted as 1) is \( kT\ln 2\) (Figure \ref{do}), in accordance with Landauer's erasure principle  (Resetting to the right can also be accomplished by first resetting to the left, followed by applying a NOT gate).

\begin{figure}
\begin{center}
{\includegraphics[scale=.5]{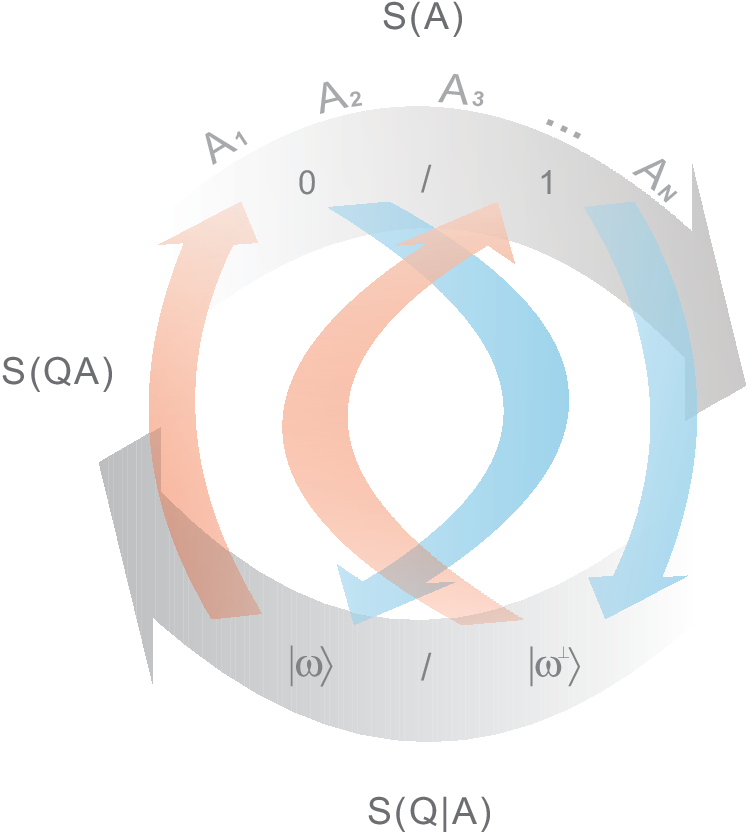}}
\end{center}
\caption{ {\bf{A recursive relation involving quantum conditional entropy.}} Using negativity in quantum conditional entropy, the relationship between \( S(Q|A) \), representing the information about \( Q \) given \( A \), and \( S(A) \), the outcome of \( A \), can be viewed recursively: based on the information from \( |\omega\rangle_Q \) or \( |\omega^{\perp}\rangle_Q \), work can be extracted and used to generate \( A \) as 0 or 1, respectively. This illustrates that, while the information about \(Q\) given \(A\) depends on the outcome of \(A\), the outcome itself arises from the information about \(Q\) with respect to \(A\).} 
\label{recur}\end{figure}

Taking into account both the causal relationship and the expression of the system \(A\)'s outcome as \(S(A) = S(QA) - S(Q|A)\) from the conditional entropy formula, the following proposition can be established:
\begin{enumerate}
\item [(L2)]  The extractable work from the conditional entropy \(S(Q|A)\) generates the outcome in \(A\).
\end{enumerate}
The recursive relationship between (L1) and (L2), as depicted in Figure \ref{recur}, demonstrates that the outcome of \(A\) originates from the information in \(Q\) relative to \(A\), influenced by the negative quantum conditional entropy. Specifically, the outcome of 0 or 1 in system \(A\) is derived from the work associated with the information of the states \(|\hat{\omega}\rangle_Q\) or \(|\hat{\omega}^{\perp}\rangle_Q\), respectively.


\section{Thermodynamics of Living Systems}

Living systems maintain order and organization by constantly using energy to counteract entropy, which naturally drives systems toward disorder. By converting energy from their environment (e.g., through metabolism), they sustain low internal entropy, enabling complex processes like growth, reproduction, and homeostasis \cite{schneider1}. In \cite{schrodinger}, the idea that life feeds on negative entropy to sustain its structure was introduced, later developed through thermodynamic models. The work on dissipative structures by Prigogine \cite{prigogine} demonstrated that living systems far from equilibrium maintain order through energy dissipation. More recent research \cite{england} suggests that living matter can spontaneously organize within energy-dissipating systems.

In the previously described protocol, the ancilla system \(A\) was initially defined as a composite of \(N\) subsystems, all starting in the state 0. Subsystem \(A_1\), which interacts with qubit \(Q\) and becomes correlated with it, registers the outcome. This orthogonal information is then copied to all \(A_i\), where \(2 \leq i \leq N\), through reversible operations. Subsequently, the subsystems \(A_{N-1}\) to \(A_1\) can be reset to their initial state of 0. Since these are known states, i.e., for \( A_N \), the reset process can be performed without energy dissipation (In fact, this procedure of copying and resetting is equivalent to successively swapping \( A_1 \) with \( A_2 \), \( A_2 \) with \( A_3 \), and so on, until \( A_N \). Again, this process is reversible and requires no energy dissipation). The information initially registered in \(A_1\) is finally encoded solely in \(A_N\), leading to a modified version of (L2): 
\begin{enumerate}
\item [(L2')] The extractable work from the conditional entropy \(S(Q|A)\) generates the outcome in \(A_N\).
\end{enumerate}
Thus, subsystem \(A_1\) functions as a measurement apparatus in the conventional sense, subsystems \(A_2\) to \(A_{N-1}\) act as intermediaries, and \(A_N\) represents the final living system that holds information about \(|\hat{\omega}\rangle_Q\) or \(|\hat{\omega}^{\perp}\rangle_Q\), corresponding to \(S(Q|A)\).

This conclusion aligns with \cite{schrodinger}, where living systems sustain themselves and reduce local entropy by consuming energy.  Notably, the recursive relation between systems \( Q \) and \( A \) arises from the negativity of \( S(Q|A) \), a distinct quantum phenomenon. Unlike classical conditional entropy, which is non-negative and doesn't involve a reversal process, this quantum feature renders the information in subsystems \( A_1 \) through \( A_{N-1} \)—including the detector and intermediate systems—irrelevant, allowing them to be reset without energy dissipation.

\section{Discussion}

In quantum information science, the probabilistic nature of outcomes stems from the principle of superposition, where a quantum system exists in a coherent combination of states until a measurement causes it to collapse into one of the possible outcomes with a defined probability. In living systems, the balance between entropy and energy plays a key role in maintaining structure and function where organisms harness energy from their environment to reduce local entropy and sustain life, much like the extractable work from quantum correlations discussed earlier. The analogy presented here posits that certain energy in biological systems parallels the useful information or extractable work from quantum correlations, suggesting a connection between the probabilistic framework of quantum systems and the thermodynamics underlying living organisms.

\end{document}